\documentclass[aip,jcp,reprint]{revtex4-1}

\usepackage{graphicx}

\usepackage[intlimits]{amsmath}
\usepackage{amsfonts}
\usepackage{amssymb}
\usepackage{mathtools}

\newcommand{\integ}{\int \!}
\newcommand{\diff}{\, \mathrm{d}}

\newcommand\addnumber{\addtocounter{equation}{1}\tag{\theequation}}

\begin{document}

\title{Accelerated weight histogram method for exploring free energy landscapes}

\author{V. Lindahl, J. Lidmar, and B. Hess}
\affiliation{Department of Theoretical Physics and Swedish e-Science Research Center, KTH Royal Institute of Technology, 10691 Stockholm, Sweden}

\date{\today}

\begin{abstract}

Calculating free energies is an important and notoriously difficult task for molecular simulations.
The rapid increase in computational power has made it possible to probe increasingly complex systems, yet
extracting accurate free energies from these simulations remains a major challenge.
Fully exploring the free energy landscape of, say, a biological macromolecule typically requires sampling large conformational changes and slow transitions. 
Often, the only feasible way to study such a system is to simulate it using an enhanced sampling method.
The accelerated weight histogram (AWH) method is a new, efficient extended ensemble sampling technique which adaptively biases the simulation to promote exploration of the free energy landscape.
The AWH method uses a probability weight histogram 
which allows for efficient free energy updates
and  results in
an easy discretization procedure.
A major advantage of the method is 
its general formulation, making it 
a powerful platform for developing further extensions and analyzing its relation to already existing methods.
Here, we demonstrate its efficiency and general applicability by calculating the potential of mean force along a reaction coordinate for both a single dimension and multiple dimensions.
We make use of a non-uniform, free energy dependent target distribution in reaction coordinate space so that computational efforts are not wasted on physically irrelevant regions.
We present numerical results for molecular dynamics simulations of lithium acetate in solution and chignolin, a 10-residue long peptide that folds into a $\beta$-hairpin. 
We further present practical guidelines for setting up and running an AWH simulation.
\end{abstract}

\maketitle 

\section{Introduction}
Free energy calculations are
 a common objective of many molecular computer simulations of chemically and biologically interesting systems.
Experiments alone only capture  details of
the thermodynamically stable end or intermediate states, which is often not sufficent for extracting detailed information about the transitions between states.
In contrast, the free energy  provides quantitative information about the available conformations and the transitions between them. 
Molecular dynamics (MD) is a valuable tool in mapping out the free energy landscape. 
However, the time step in MD simulations is limited to a few femtoseconds, whereas 
events of interest are often  characterized by timescales of micro- or milliseconds. 
Canonical MD simulations can reach microseconds, but even that is often not enough to obtain sufficient statistics of the relevant transitions. The underlying issue is that macromolecules are often characterized by high free energy barriers. 
Since transition times scale exponentially with the barrier height, transitions across such barriers are difficult-to-sample, rare events.

One popular method for improving the sampling efficiency in free energy calculations is steered MD, in which the system is pushed forward along a reaction coordinate, followed by umbrella sampling \cite{Torrie:1977hs} along the obtained path. For systems with a narrow transition valley this may work well, provided a good reaction coordinate is chosen. 
However, due to the non-equilibrium nature of the steering and the high forces involved, one risks pushing the system into improbable and non-representative states.
If the free energy landscape is rough, 
the subsequent umbrella sampling will not be able to efficiently relax the system.
Furthermore, this procedure only samples a single pathway (unless run multiple times), whereas there could be multiple pathways available for the same transition. 
A more flexible and general approach is called for.

Extended, or generalized, ensemble methods\cite{Iba:2001hy, Mitsutake:2001kf} is a collective name for a wide variety of sampling techniques 
in which the original ensemble is modified in order to overcome the limitations of conventional sampling methods.
Because of the inherent flexibility of the idea, it has been used in many different applications of physics, including simulations of spin models\cite{Berg:1992uh, Lidmar:1375762}, nucleation \cite{Chen:2001vu} and protein folding \cite{Hansmann:1997hi}, just to name a few.

One extended ensemble approach is to promote a system parameter, e.g. temperature, to be a dynamical variable and
within a single simulation perform a biased random walk in parameter space while maintaining a canonical distribution at each fixed parameter value.
A well-known example is simulated tempering \cite{Marinari:236598,Lyubartsev:1992th}, in which the system is randomly heated up or cooled down according to the joint coordinate-temperature distribution, increasing the chance of crossing high energy barriers.
The biasing in parameter space is determined by assigning to each parameter value a probability weight factor such that the parameter space gets sampled according to a certain target distribution of choice, often simply chosen uniform.
The particular set of weights that give rise to the specified target distribution is related to the initially unknown free energy landscape of parameter space through a simple relation. Thus, finding the correct weights is a major challenge since it amounts to calculating the free energy.

A general strategy is to, starting from an initial guess, adaptively refine the weights using the simulation history. 
Adaptive biasing procedures have been developed both for Monte-Carlo (MC) simulations \cite{Kumar:1996tk, Wang:2001eb,Belardinelli:2007ce} 
in the  multicanonical ensemble \cite{Berg:1992uh},
in which the 
canonical weights are modified to obtain a uniform distribution of energies, as well as in the context of MD by modifying the potential energy (or force) along the system trajectory \cite{Bartels:1997wl,Huber:1994db,Grubmuller:1995in,Laio:576913,Barducci:2008ua,Darve:2001ce,Dickson:2010dh,Kim:2006ev}. 
Many of these methods are closely related and in some cases functionally equivalent \cite{Junghans:2014tz}, but since their development has largely taken place in parallel, relatively little effort has been devoted to investigating their exact relations.
Nonetheless, it is often possible to adapt these methods to the extended ensemble formalism, making this an attractive platform for unifying knowledge and working on further developments.

In this paper we study in detail the 
 recently introduced accelerated weight histogram (AWH) method \cite{Lidmar:1375762}, an adaptively biasing, extended ensemble method equipped with several advantageous features, including:
(i) allowing for large transitions in parameter space by  using a Gibbs sampler,
(ii) using a probability weight histogram 
to efficiently adapt future bias based on the transition history and
 which further makes the binning procedure simple,
(iii) being formulated within a very general extended ensemble framework which 
makes the method highly customizable and applicable to a wide variety of problems. 

Here, we propose two highly useful extensions to the AWH method.
First, we show how to calculate the potential of mean force (PMF), the free energy along a reaction coordinate. 
Second, we explore non-uniform, free energy dependent target distributions in parameter space for which sampling of irrelevant regions of phase space is automatically avoided. Numerical results are presented for MD simulations of relatively simple chemical and biological molecular systems which will serve as benchmarks for future, more complex applications. Furthermore, we study the input parameters of the method in detail and provide guidelines for applying the method in practice.

The AWH method in its general form is described in section \ref{sec:basicalgorithm}.
In section \ref{sec:pmf}, we build on the basic algorithm  by providing a procedure for calculating the PMF.
The choice of target distribution is discussed in
section \ref{sec:histdeptargetdistr}, where we also present a couple of concrete alternatives.
 In section \ref{sec:n}, we investigate another important input parameter, namely the effective number of samples, which in the AWH method sets the bias update size. We propose how to initialize and update the simulation in order to obtain a robust and efficient method. 
  Finally, in section \ref{sec:applications} we discuss the practical aspects of setting up an AWH simulation and demonstrate the strengths of the method for two molecular test systems: solvated lithum acetate and chignolin, a 10-residue peptide. We conclude in section \ref{sec:conclusion}.

\section{The accelerated weight histogram method}

\begin{table}
\setlength{\tabcolsep}{8pt}
\footnotesize
\begin{tabular}{ l | l}
\hline
  $x, \mathcal{X}$ &  configuration, configuration space \\
  $\lambda, \Lambda$ & extended parameter, parameter space \\
  $\xi(x)$ & reaction coordinate \\ 
  $f(\lambda)$ & estimate of the true free energy $F(\lambda)$ \\
  $\phi(\xi)$ & estimate of the true PMF $\Phi(\xi)$ \\
  $g(\lambda)$ & biasing function \\
  $\rho(\lambda)$ & target distribution along $\lambda$ \\
  $\omega(\lambda)$ & $\lambda$ transition probability distribution \\
  $N$ & effective number of $\lambda$ samples \\ 
  $W(\lambda)$ & reference weight histogram \\
  $n_\Lambda$ &  number of $\lambda$ samples per update \\ 
  $ \Delta t_\Lambda$ & time between $\lambda$ samples \\ 
  $S$ & number of collected $\lambda$ samples \\ 
  $\kappa$ & umbrella potential force constant \\ 
  \hline
\end{tabular}
\caption{Summary of the main AWH variables and their meaning.}\label{tab:notation}
\end{table}

\subsection{\label{sec:basicalgorithm} The basic algorithm}
We consider a system of particles described by configurations $x\in\mathcal{X}$ and
a system parameter $\lambda$, possibly $d$-dimensional (see table \ref{tab:notation} for a summarizing table of the notation we will be using). For instance, 
$\lambda$ could be a thermodynamic state  parameter such as temperature or pressure.
$\lambda$ may be of continuous nature but for all practical purposes it can be considered discrete.
We assume that the equilibrium probability distribution of the system is given by 
$\pi(x;\lambda) = e^{-E(x;\lambda) + F(\lambda)}$, where
$F(\lambda)$ is the dimensionless free energy  along $\lambda$ 
(i.e.\ the free energy scaled by $\beta = 1/k_B T$) and is defined by 
$F(\lambda) = -\ln \integ e^{-E(x;\lambda)}\diff x$. 

We now assume that our goal is to explore the free energy landscape $F(\lambda$). 
In the extended ensemble, $\lambda$ is
promoted to be a dynamic variable alongside $x$, and is allowed to take on a range of values, $\lambda\in\Lambda$.
The extended ensemble 
is thus described by the joint distribution 
$P(x,\lambda) = \frac{1}{\mathcal{Z}} e^{-E(x,\lambda) + g(\lambda)}$,
where $g(\lambda)$ is a biasing function
that is tuned during the simulation to obtain a certain user-specified target distribution $\rho(\lambda)$. 
The actually observed marginal distribution for $\lambda$, $P(\lambda)$, is related to the unknown free energy $F(\lambda)$ by
\begin{align*}
 P(\lambda) &= \integ P(x,\lambda)\diff x 
			= \integ \frac{1}{\mathcal{Z}} e^{-E(x,\lambda) + g(\lambda)} \diff x \\
 &=\frac{1}{\mathcal{Z}} \, e^{-F(\lambda) + g(\lambda)},
\addnumber{\label{eq:plambda}}
\end{align*}
which is generally \emph{not} equal to the target distribution $\rho(\lambda)$ unless $g(\lambda)$ has been tuned to balance out $F(\lambda)$ correctly. 
To achieve this in the simulation, $g(\lambda)$ is chosen consistently with equation \eqref{eq:plambda} by substituting $P(\lambda)$ with $\rho(\lambda)$ and 
 $F(\lambda)$ with $f(\lambda)$, our best estimate of the free energy, yielding
\begin{equation}
g(\lambda) = f(\lambda) + \ln \rho(\lambda),\label{eq:gopt}
\end{equation}
where the omitted constant is not of importance for free energy differences. $f(\lambda)$ is initialized, e.g.\;by guessing, and is then iteratively refined based on the sampling history, as described below.

$P(x,\lambda)$ is sampled by
 performing $n_{\mathcal{X}}$ updates of $x$ at fixed $\lambda$, using standard MD or MC, alternated by an update of $\lambda$ at fixed $x$. 
In the AWH method, $\lambda$ is updated using a Gibbs sampler. That is, a new $\lambda$ is chosen according to the 
probability distribution
\begin{equation}
\omega(\lambda|x) 
= P(\lambda|x) 
= \frac{1}{Z_\omega}e^{-E(x,\lambda) + g(\lambda)}.
\label{eq:omegalambdax}
\end{equation}
After having collected $n_\Lambda$ samples of $\lambda$, $f(\lambda)$ is updated.
Since $P(\lambda) = \integ \omega(\lambda|x) P(x) \diff x$,
the sum of transition probabilities 
$\sum_{i=1}^{n_\Lambda} \omega^i(\lambda)
=:n_\Lambda \bar\omega(\lambda)$ 
can be used to estimate the current discrepancy between  $P(\lambda)$ and the desired target distribution $\rho(\lambda)$.
In the AWH method, $n_\Lambda\bar{\omega}$ is seen as a fluctuation on top of a perfectly distributed reference weight histogram $W(\lambda)$ containing $N$ effective number of samples, i.e.
\begin{equation*}
W(\lambda) = N\rho(\lambda).
\end{equation*}
Equations \eqref{eq:plambda} and \eqref{eq:gopt} then imply an update 
$f(\lambda)\leftarrow f(\lambda) + \Delta f(\lambda)$, where 
\begin{align*}
\Delta f(\lambda) 
&=
-\ln\left(\frac{W_{\text{fluct}}}{W_{\text{target}}}\right)
=
-\ln\left(\frac{N\rho(\lambda) + n_\Lambda\bar{\omega}(\lambda)}
{N\rho(\lambda) + n_\Lambda\rho(\lambda)}
\right)\\
&=-\ln
\left(
1 + \frac{n_\Lambda}{N}\frac{\bar{\omega}(\lambda)}{\rho(\lambda)} 
\right) + \text{constant}. 
\addnumber{\label{eq:df}} 
\end{align*}
The constant is in principle not of importance but for numerical reasons it should be included in the implementation. 
Next, the bias $g(\lambda)$ is updated in a consistent manner by applying equation \eqref{eq:gopt} for the newly updated  $f(\lambda)$.
For $\rho$ uniform we recover, up to a constant, $f=g$ as in \cite{Lidmar:1375762}, where non-uniform $\rho$ was not explicitly treated.
Finally, the effective number of samples $N$ is updated by 
$N \leftarrow N + \Delta N$, where $\Delta N = n_\Lambda$.
 We note that this update
is the normal running condition of the algorithm. However, in principle, the updates of  $N$ can be chosen more generally, e.g. in the initial stages of the algorithm or to improve a poorly converging run (section \ref{sec:updatingN}), or in more exotic method setups (section \ref{sec:effectivetempdistr}).

From  equation \eqref{eq:df} we see that $\Delta f$ decreases as $\sim 1/N$ for large $N$, allowing for increasingly fine resolution of the free energy to be probed.
 The AWH method continues iteratively in this way e.g. for a fixed number of steps.

The main distinguishing feature of the AWH method is the use of a Gibbs sampler in $\lambda$,
which enhances mixing of $\lambda$ relative to nearest neighbor sampling  
\cite{Chodera:2011ia},
in combination with  updates of $f$ that efficiently make use of all the available sampling history, including those transitions that had a probability to take place but did not. 
With this choice of updates
 discretizing $\lambda$ becomes trivial and non-critical for the efficiency of the method as long as the point spacing is dense enough to make transitions likely to occur.
This is a clear advantage over other methods  for which the discretization or binning procedure can be not only tedious but also critical for the performance. For instance, in umbrella sampling one system has to be equilibrated and run in each umbrella. 
Another major advantage of the AWH method is that it allows the system to explore multiple pathways $x(t)$ when traversing $\Lambda$ so that the results do not critically depend on the quality of the initial configuration. 

In addition, being a histogram-based method, the AWH method is inherently straightforward to parallelize. 
One particularly simple and implementation-friendly scheme is to simultaneously carry out multiple simulations, each generating samples along its own independent trajectory but sharing the same biasing weights. 
An alternative approach is to run multiple non-communicating simulations in parallel and combine them to a final estimate $\bar{F}$ in the end 
 as described in \cite{Lidmar:1375762}. 
Although letting the replicas communicate may speed up convergence, at least initially, the advantage of this approach is that the statistical error of $\bar{F}$ can be calculated using standard jackknife statistics \cite{Berg:2004vc}.

\subsection{\label{sec:pmf}Free energy along a reaction coordinate}
In many real-world applications we are not interested in the free energy as a function of a system \emph{parameter} $\lambda$.
In systems with many degrees of freedom it might for instance not help to heat up the system since  the accessible part of phase space would increase drastically, potentially hampering sampling of important (low energy) states.
In such cases it may be favorable to be more selective and incorporate prior knowledge into a, possibly multidimensional, reaction coordinate $\xi(x)$.
As a simple example, $\xi$ could be a distance or an angle that is known to  be involved in the transition of interest. 
The corresponding free energy $\Phi(\xi)$, or the potential of mean force (PMF), is defined by 
\begin{equation}
\Phi(\xi) = -\ln{\integ \pi_0(x)\delta(\xi-\xi(x)) \diff x},
\end{equation}
where $\pi_0(x)$ is the equilibrium distribution of the system.
In the reaction coordinate case, it is not possible to move $\xi(x)$ independently of $x$ or vice-versa. Nonetheless, we can calculate $\Phi(\xi)$ with the AWH method by coupling the system to a set of harmonic potentials, or umbrellas, 
\begin{equation}
Q_\kappa(\xi,\lambda) = \frac{\kappa}{2}(\xi-\lambda)^2,
\label{eq:qkappa}
\end{equation}
with centers at $\lambda\in\Lambda$.
We can make jumps between the umbrella centers $\lambda$ independently of $x$.
 The dynamics of $\xi$  effectively follows that of $\lambda$ if the force constant $\kappa$ is chosen large enough. This setup corresponds to pulling $\xi$ towards the umbrella center $\lambda$ using a harmonic spring.
 
The algorithm will however not estimate $\Phi(\xi)$ directly but rather the free energy $F(\lambda)$ of 
 the original ensemble modified by the umbrellas,
\begin{align*}
 e^{-F(\lambda)}
 &= \integ e^{-E(x;\lambda)} \diff x
 = \integ  e^{-Q_\kappa(\xi(x),\lambda)} \pi_0(x) \diff x\\
 &= \integ e^{-Q_\kappa(\xi,\lambda)} e^{-\Phi(\xi)}  \diff\xi. \addnumber{\label{eq:convolution}}
\end{align*}
For large $\kappa$, $F\approx\Phi$, while for smaller $\kappa$, $F$ will appear increasingly smeared relative to $\Phi$. 
%
\begin{figure}[tbp!] 
\includegraphics[width=0.38\textwidth]{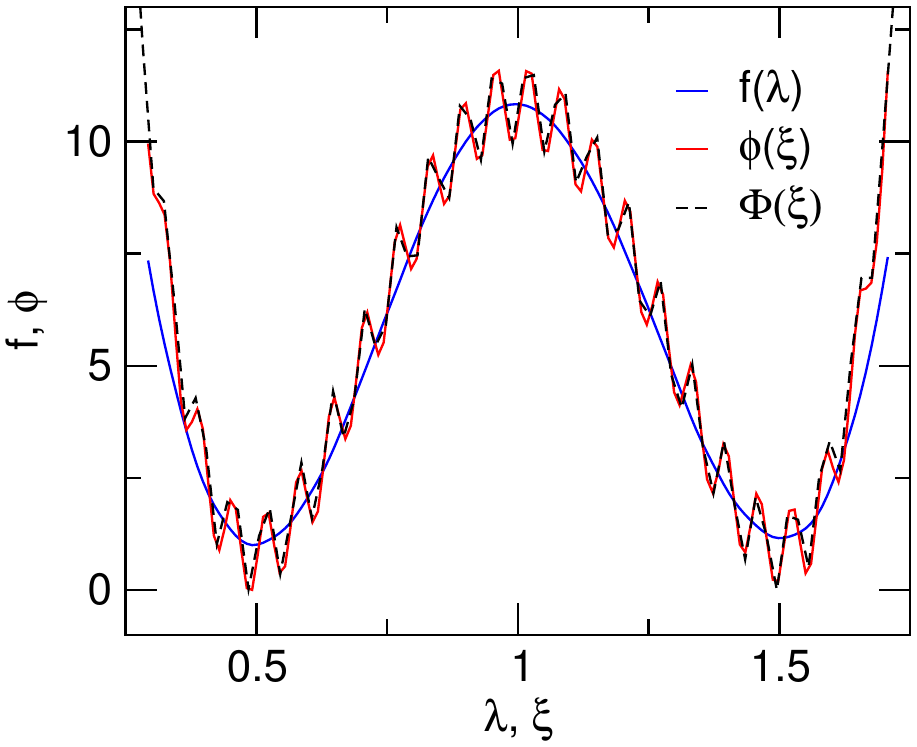}
\caption{\label{fig:deconv} Extracting the estimated PMF $\phi(\xi)$ from the convoluted free energy estimate $f(\lambda)$ using equation \eqref{eq:deconvavg} for a Brownian particle in the potential $\Phi(\xi) = 80 (2(\xi-1)^4-(\xi-1)^2) + \sin(100\xi)$. The PMF is seen to be recovered with full resolution.}
\end{figure}
Although $\Phi$ can, in principle, be recovered by directly solving eq.\,\eqref{eq:convolution}, this is
unfortunately a rather ill-conditioned problem which may give rise to numerical inaccuracies. 
Here, we propose instead to deconvolute equation \eqref{eq:convolution} on the fly by making use of the collected samples of $\xi$.
Using the fact that the marginal distribution is
\begin{align*}
P(\xi) 
&= \sum_\lambda \integ P(x,\lambda) \delta(\xi - \xi(x))\diff x\\
&= \sum_\lambda \frac{1}{\mathcal{Z}} \, e^{ 
-Q_\kappa(\xi,\lambda) - \Phi(\xi) + g(\lambda)},
\end{align*}
$\Phi$ can be solved for as
\begin{equation}
e^{-\Phi(\xi)} = 
\mathcal{Z} e^{\gamma(\xi)} P(\xi),
\label{eq:deconv}
\end{equation}
where $e^{-\gamma(\xi)} = 
\sum_\lambda e^{g(\lambda) - Q_\kappa(\xi,\lambda)}$.
Direct application of equation \eqref{eq:deconv} is complicated by the fact that the bias $g$, and hence the entire ensemble is being updated at each iteration. 
First of all, the unknown normalization constant
$\mathcal{Z} = \sum_\lambda e^{-F(\lambda) + g(\lambda)}$
changes with each update, which seemingly makes the averaging of eq.\,\eqref{eq:deconv} over different iterations problematic. Nonetheless, $\mathcal{Z}$ tends to an irrelevant constant in  the later stages of the algorithm and can in our experience safely be ignored. 
Second, in order to ensure that $\phi$, the estimate of $\Phi$, is updated consistently with $f$, the histogram $e^{-\phi(\xi)}$ should be rescaled with a factor $a=(N+\Delta N)/(N+n_\Lambda)$ after each update such that it grows at the same rate as $W(\lambda)$ (which determines the size  of $\Delta f$). 
For the standard update $\Delta N = n_\Lambda$ this reduces to a trivial scaling of 1. However, in the initial stages of the algorithm often a more heuristic update of $N$ will be applied (see section \ref{sec:updatingN}). 
For instance, if $N$ is temporarily kept constant, $\Delta N=0$, the downscaling $N/(N+n_\Lambda)$ will ensure that fluctuations in both $f$ and $\phi$ are 
kept at a constant overall magnitude $\sim 1/N$.
To summarize, we propose calculating the PMF estimate $\phi$ using the time average
\begin{equation}
e^{-\phi(\xi)} = 
\frac{\langle a(t) e^{\gamma(\xi,t)}\delta(\xi-\xi(t)\rangle_t}
{\langle a(t) e^{\gamma(\xi,t)} \rangle_t},
\label{eq:deconvavg}
\end{equation}
where $\delta$ is a binning function. 

Figure \ref{fig:deconv} demonstrates how the PMF $\phi(\xi)$ can be extracted from the free energy $f(\lambda)$ using the deconvolution procedure of equation \eqref{eq:deconvavg}.
Our simple test system is a Brownian particle at $k_BT=1$ moving in a "rugged" double-well potential $\Phi(\xi) = 80 (2(\xi-1)^4-(\xi-1)^2) + \sin(100\xi)$. We use umbrellas of curvature $\kappa=1024$, evenly spaced in $\lambda\in (1\pm {1}/{\sqrt{2}})$. In the figure, we see that the fine-structure of the potential is smeared out in $f(\lambda)$, but is fully recovered in $\phi(\xi)$.
The high resolution is made possible by the extra information added by the sampling in $\xi$. 
We were not able to recover this fine structure of the potential using the standard Richardson-Lucy deconvolution algorithm\cite{Richardson:1972vb,Lucy:1974uy}, which in addition is known to be sensitive to the number of performed iterations. Another advantage to our scheme is that no post-processing is needed.

\subsection{\label{sec:histdeptargetdistr} Choosing the target distribution $\rho$}
The target distribution $\rho(\lambda)$ is often simply chosen to be uniform since this increases the probability of crossing high free energy barriers. In addition, $\rho$ can of course take on any explicit dependence on $\lambda$,
e.g  if there is prior knowledge available about which regions of $\Lambda$ should be explored more or less.
For one-dimensional reaction coordinates there has also been promising developments \cite{Singh:2011jj, Tian:2014ec} in diffusion-optimized biasing methods where the target distribution becomes a function of the position-dependent diffusion coefficient in order to minimize the passage time across $\Lambda$.  

Still, for some complex systems,
especially in the multidimensional case,
 it is often difficult to \emph{a priori} define the sampling region $\Lambda$ such that all important states are accessible 
but the improbable, high free energy regions excluded. 
Inclusion of irrelevant regions may lead  to poor convergence and even irreversible damage, such as breaking of important bonds.
We address this issue by adding a simple extension to the basic AWH
algorithm; namely, we let the target distribution $\rho(\lambda)$ be a
decreasing function of the free energy $F(\lambda)$ such that regions with too high free energy are avoided. 
Seeing that $F$ is unknown however, we have to resort to using our best estimate, $f$.
Furthermore, since $f$ is constantly being updated we must, after updating $f$ and before updating the bias function $g$, update $\rho$.
Below, we present two target distributions that both deal with this $\Lambda$ boundary problem.

\subsubsection{\label{sec:targetdistrcutoff} Target distribution with free energy cutoff}
One way of specifying $\rho(\lambda,t)$ without risking to push the system into irrelevant regions
is to set it to a fixed function $\rho_0(\lambda)$ for $\lambda$ points with $f(\lambda)$ below a given cutoff $C$, and let it decay exponentially  with $f$ otherwise. Specifically, at each update time we set $\rho$ according to
\begin{equation}
\rho(\lambda) = 
\begin{cases}
\frac{1}{Z} \rho_0(\lambda) &\mbox{if } f(\lambda) \leq f_C \\
\frac{1}{Z} \rho_0(\lambda) e^{-(f(\lambda)-f_C)} &\mbox{if } f(\lambda) > f_C
\end{cases},\label{eq:pic}
\end{equation}
where the free energy cutoff $f_C$ is measured relative to the global minimum, i.e. $f_C = \min_{\lambda}f(\lambda) + C$.

\subsubsection{\label{sec:effectivetempdistr} Target distribution with effective temperature}
 A more continuous fashion of focusing sampling to low free energy regions 
 would be to let $\rho$ be a Boltzmann distribution with an effectively raised temperature, $T+\Delta T$. 
The most straightforward way to achieve this in the AWH framework is to set  $\rho\propto e^{-\frac{T}{T+\Delta T}f}$ and update the target distribution as
$-\Delta \ln \rho = T/(T+\Delta T)\cdot \Delta f$.
 
 We can also take inspiration from well-tempered metadynamics \cite{Barducci:2008ua}, a popular adaptive biasing method 
in which the  distribution along the reaction coordinate $\xi$  converges exactly to a tempered Boltzmann distribution,
$P_{\infty}(\xi) \propto e^{-\frac{T}{T+\Delta T}\Phi(\xi)}$.
 In well-tempered metadynamics, repulsive potentials, usually Gaussian, of a certain height are dropped along the reaction coordinate trajectory $\xi(t)$, which forces the system to explore new regions of phase space.
The deposit height is set to decrease as $\sim e^{-\frac{V(\xi,t)}{\Delta T}}$
(i.e.\, is dependent on $\xi$), where $V(\xi,t)$ is the total biasing potential. 

We can recover a setup very similar to well-tempered metadynamics as a special case  of the AWH method
 by defining the target distribution $\rho$ as an explicit function of the sampling history. We leave the details of this variant to the appendix \ref{app:efftempdistr}. The main difference of the AWH formulation is the use of the extended ensemble: $\xi$ is not directly biased, but indirectly via $\lambda$. Furthermore, the shape of $\Delta f$ is not constant, but determined dynamically by equation \eqref{eq:df}. 
 
 It is unclear if this type of setup  would actually offer any advantages to using a fixed target distribution (possibly combined with a free energy cutoff). There may even be a downside to letting $\rho$  be completely determined by the sampling history;  validating the simulation status by comparing the empirical distribution with the ideal one, $\rho$, becomes meaningless. 

\subsection{\label{sec:n} The effective number of samples, $N$}
The effective number of samples $N$ is an important parameter in the AWH method since it determines the overall size of the update, $\Delta f \sim 1/N$ (equation \eqref{eq:df}). 
Small $N$ values (large $\Delta f$) are associated with high transition rates and large fluctuations in $f$, 
while larger $N$ values (small $\Delta f$) yield slower dynamics and allow for a more accurate free energy estimate.
Because of its relation to the free energy error, the initializing and updating of $N$ deserves some extra attention. 
Nonetheless, as we will see further on in this section, with our proposed $N(t)$ protocol, the performance of the method becomes fairly insensitive to the initial setting of $N$.

\subsubsection{The error and the saturated error,  $\varepsilon(t)$ and $\varepsilon_{\text{sat}}$}
In this paper we use the error measure 
\begin{equation}
\varepsilon^2(t) =
\frac{1}{|\Lambda|}
\sum_\lambda
\left\langle
\left(f(\lambda,t) - F(\lambda)\right)^2\right\rangle \label{eq:epsdef},
\end{equation}
where $|\Lambda|$ is the number of points in $\Lambda$ and $\langle\cdot\rangle$ denotes statistical averaging
over independent simulations.
When $F$ is unknown, we use $\langle f\rangle$ instead (decreasing the statistical degrees of freedom by 1).
When averaging over several simulations, we first align each free energy profile such that 
$\frac{1}{|\Lambda|}\sum_\lambda f(\lambda) - F(\lambda) = 0$.
For the PMF error we simply make the replacements $f\mapsto\phi$, $F\mapsto\Phi$, and $|\Lambda|\mapsto|\mathcal{X}|$, the number of $\xi$ bins.
Since our free energy variables are defined as dimensionless, their errors are also dimensionless. Units of energy are obtained by scaling with $k_BT$.

Consider now a simulation where $N$ is kept constant.  Then
$f$ will only be refined up to a certain level before the error  $\varepsilon(t)$ saturates, $\varepsilon(t) \to \varepsilon_{\text{sat}}$,
$t\to \infty$, where 
$\varepsilon_\text{sat} = \varepsilon_\text{sat}(N)$. 
When letting $N$ increase with time however, $N=N(t)$,
$\varepsilon_{\text{sat}}(N(t))$ will decrease with time and
the actual error
$\varepsilon(t)$ 
will only stay close to saturation if $N(t)$ grows at a slow enough rate.
For convenience we introduce 
\begin{equation}
\eta(t) = \frac{\varepsilon(t)}{\varepsilon_{\text{sat}}(N(t))}, 
\label{eq:eta}
\end{equation}
 which is a measure of how far the error is from saturation.  Holding $N$ constant, $\eta(t)\to 1$.
If $N$ increases too rapidly, $\varepsilon(t)$ cannot follow $\varepsilon_{\text{sat}}(N(t))$ and $\eta(t)$ grows.  

\subsubsection{\label{sec:initializingn} Initializing $N$}
The initial effective number of samples, $N^0$, should ideally
reflect the quality of the initial guess of the free energy, $f(\lambda)$, which
typically will be quite inaccurate. Thus, given a rough estimate of the initial error
 in $f$, $\varepsilon_0=\varepsilon(0)$, we would like to estimate an appropriate $N^0$.
Obviously, $\varepsilon_0$ is not known initially but can e.g. be estimated based on a guess of typical barrier heights. 
We further assume for the time being that also $\varepsilon_\text{sat}$ can roughly be estimated.

Based on equation \eqref{eq:eta}, it is natural to 
 aim for an $N^0$ for which $\eta(0)=\eta^0\gtrsim 1$, such that the initial error is close to the saturated one ($\eta \approx 1$) but still tends to decrease ($\eta > 1$).
In our experience however (see section \ref{sec:langevin}), there are benefits to choosing $N^0$ on the smaller rather than the bigger side, meaning
\begin{equation}
\eta^0 
= \frac{\varepsilon_0}{\varepsilon_{\text{sat}}(N^0)}
\lesssim 1.
\label{eq:eps0epssatN0}
\end{equation}
Since 
$\eta(t)$ tends to 1,
 $\varepsilon(t)$ might increase initially.
Still, as long as one takes care to not drive the system out of equilibrium by choosing $N^0$ extremely small, it is more useful to see one transition than none at all, which one risks by setting $N^0$ too large.  

\subsubsection{\label{sec:updatingN} Updating $N$}
As was assumed in the description of the basic algorithm (section \ref{sec:basicalgorithm}), the effective number of samples $N$ most naturally grows with the collected number of $\lambda$ samples 
$S\propto t$, where $t$ is the simulation time. 
That is, 
$N(t) = N_{\text{ref}}(t) = N^0 + S(t) \sim t$.
In addition, both theoretical and numerical studies of adaptive biasing methods support that possibly optimal convergence, $\varepsilon^2(t) \sim 1/t$,
 is obtained by asymptotically letting the bias update size decay as $\sim 1/t$ \ \cite{Barducci:2008ua, Belardinelli:2007gs,Zhou:2008cm,Belardinelli:2014ur}. 
 This is consistent with $\Delta f\sim 1/N \sim 1/S \propto 1/t$.

On the other hand, a more conservative scheme, where $N$ increases slower than $t$, will increase the robustness of the method and is useful in the early stages, before the available phase space has been sufficiently explored.
The initial stages of the algorithm are often characterized by large errors and filling up of deep free energy wells, and samples tend to be highly correlated.
The basic assumption that the collected samples follow eq.\,\eqref{eq:plambda} is then inaccurate.
In this transient regime, experience suggests that initially one should let $N(t)$ follow a more heuristic updating protocol rather than $N(t)\sim S(t)$.
In approaching this issue, there are two questions to address: what evolution should $N(t)$ follow initially, and at what $N=N_{\text{exit}}$ should $N(t)\sim S(t)$ start?

The well-known Wang-Landau updating scheme \cite{Wang:2001eb} suggests an answer for the first question.
In Wang-Landau, the update size is kept constant until the histogram of visits is sufficiently flat at which point the update size is halved. This process is repeated, e.g. until the update size is smaller than some tolerance value. This strategy has proven robust and efficient at reducing the initially large errors, but is also known to fail to 
converge asymptotically, since the errors saturate at a finite value, $\varepsilon(t)\to\varepsilon_{\infty}>0$.

The second question is dealt with in 
the Wang-Landau-based method proposed in \cite{Belardinelli:2007ce}, which accommodates for
both the desired transient and the asymptotic behavior by dividing the algorithm into two stages: an initial Wang-Landau stage,  followed by a final $1/t$ stage.
The method interpolates between the two stages by exiting from the initial stage as soon as the update size has decreased to $\leq 1/t$, after which the update size is kept at $1/t$ for the rest of the simulation.
An attractive feature of this scheme is that it picks the exit time in a dynamic and automatic manner.

\begin{figure}[tbp!]
	\includegraphics[width=0.4\textwidth]{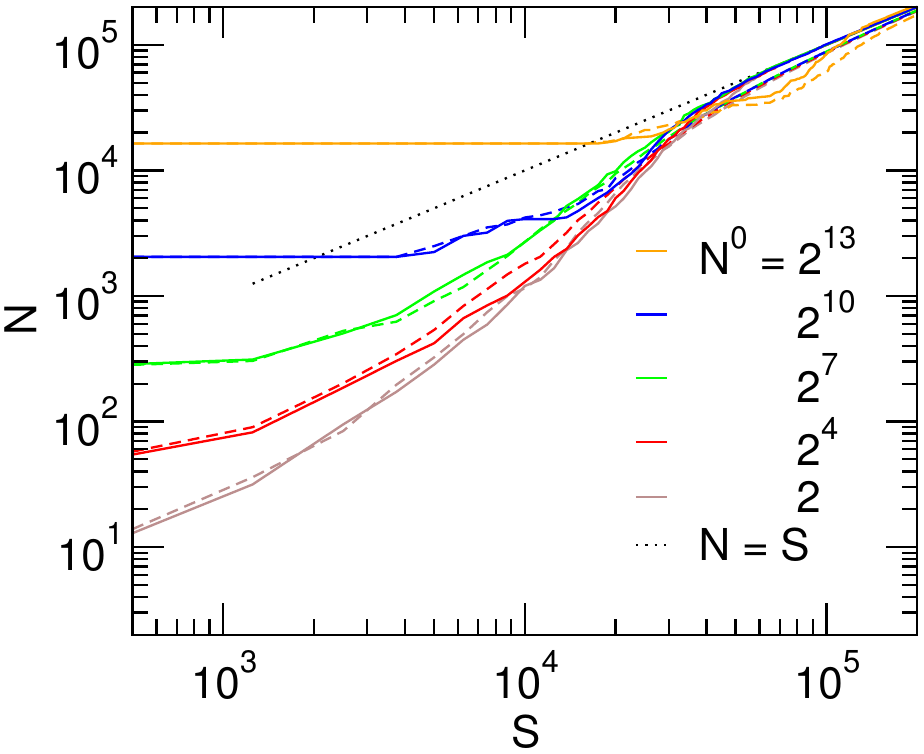}	
	\caption{\label{fig:nt} Typical evolution of the effective number of samples $N$ as a function of the number of collected samples $S\propto t$, for a range of $N^0$ values. Solid lines:  exit criterion $N \geq N_{\text{ref}}$. Dashed: $\diff N/\diff S \geq \diff N_{\text{ref}}/\diff S$. Each curve is obtained by averaging over 32 runs of a Brownian particle in a double well potential.}
\end{figure}

We can naturally adapt the WL $1/t$ approach to the AWH language. In the initial stage, $N$ is kept constant ($\Delta N = 0$) until a certain covering criterion (e.g.\, visit all of $\Lambda$) is met, triggering a doubling of $N$ ($\Delta N = N$), which leads to exponential growth initially. 
The exit occurs when 
$N(t) \geq  N_{\text{ref}}(t) = N^0 +S(t)$.
After exiting, $N$ grows linearly with time, $N(t)=N_{\text{ref}}(t) \sim t$. 

We obtain an alternative, but in practice similar, 
slope-based interpolation method by going from the initial to the final stage when the exponential growth of $N$ exceeds the linear one of $N_{\text{ref}}$, $\diff N/\diff S \geq \diff N_{\text{ref}}/\diff S$. 
This implies that the exit occurs when $N$ grows larger than 
 the number of samples collected during the most recent covering of $\Lambda$.
This can be interpreted as effectively having diffused across $\Lambda$ at least once.


Figure \ref{fig:nt} shows typical behavior of $N$ as a function of $S$ for both the $N$-based and the $\diff N/\diff S$-based types of exit criteria for a range of initial $N^0$ values in a simple test case: a Brownian particle at $k_BT=1$ moving in a one-dimensional double-well potential,
 $\Phi(\xi) = 80 (2(\xi-1)^4-(\xi-1)^2)$.
 We see that the two exit criteria in practice yield very similar evolution of $N$ and that in each case the lower values of $N^0$ on average exit to the linear stage roughly around one value $N=N_{\text{exit}}$.
The only exception occurs for the curve with a value of $N^0 \gtrsim N_{\text{exit}}$ which displays different behavior.
%

We note that doubling $N$, as we have proposed here, is in principle a dangerous operation since it corresponds to scaling up the data by a factor of 2. However, because of the form of the presented exit conditions, the average growth of $N$ will never exceed that of the "natural" sampling rate of $N_{\text{ref}}$.

In addition to the "artificial" control of $N$ initially, for certain runs it may later be advantageous to suddenly decrease $N$, i.e.\ increase the update size $\Delta f$, in order to help push the system out of potential sampling traps\cite{Poulain:2006ka}. 
To detect such situations we suggest to keep a record of the
accumulated and normalized histogram of transition weights, $\bar{\omega}_\text{tot}(\lambda)$, since its fluctuations should decay as $\sim 1/N$ asymptotically \cite{Zhou:2008cm, Swetnam:2010dz}. 
If a dramatic change of the fluctuations is observed the effective number of samples $N$ should be decreased, e.g.\ using
$N\leftarrow N\sum_\lambda \min\left(\bar{\omega}_\text{tot}(\lambda),\rho(\lambda)\right)$.
However, if the accumulated weight histogram repeatedly displays anomalous behavior, this may actually be an indication of a poorly chosen reaction coordinate.

\subsubsection{\label{sec:langevin} Test case: Langevin dynamics}
We now test the performance of our proposed initialization and update protocol for $N$ in the special, but illustrative, case of Langevin dynamics. 
First, to validate choosing $N^0$ based on equation \eqref{eq:eps0epssatN0}, we need an explicit formula for the saturated error. 
In the context of constant update size metadynamics, this has previously been derived \cite{Bussi:2006gg}: 
\begin{equation} 
\varepsilon^2_\text{sat} =
\frac{L_\xi^2 \beta w}{\tau_G D_\xi}
\left(\frac{\sigma_\xi\sqrt{2\pi}}{L_\xi}\right)^d
\zeta(\sigma_\xi/L_\xi),
\label{eq:eps2satmtd}
\end{equation}
 where $\varepsilon^2_{\text{sat}}$ is dimensionless, $L_\xi$ is the side of the $d$-dimensional cubic domain, 
$D_\xi$ is the diffusion coefficent (when $d>1$ a trace over the diffusion tensor is implied), and
 $w$ and $\sigma_\xi$ are the Gaussian height and width, which are deposited at time intervals of $\tau_G$. 
  The geometric factor $\zeta(\sigma_\xi/L_\xi) 
  =\sum_{k \neq 0} e^{-\frac{1}{2} \pi^2 k^2 \sigma_\xi^2/L_\xi^2}/\pi^2 k^2$, where $k\in \mathbb{N}^d$, increases with the number of dimensions and decreases with $\sigma_\xi/L_\xi$ as can be seen from its definition.
\begin{figure}[tbp!]
		\includegraphics[width=0.4\textwidth]{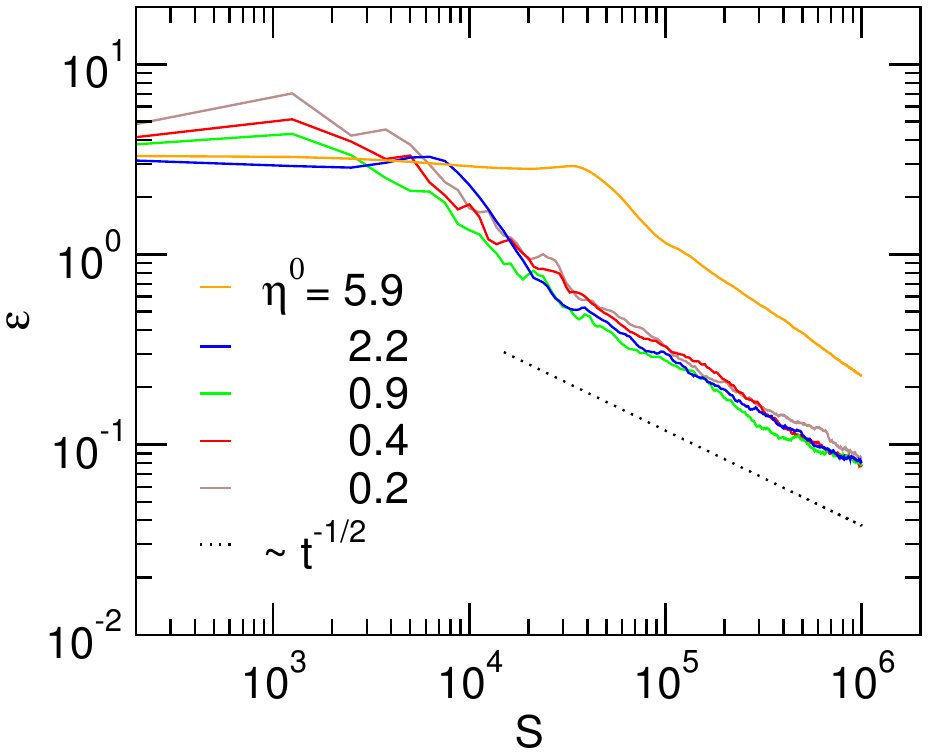}
		\includegraphics[width=0.4\textwidth]{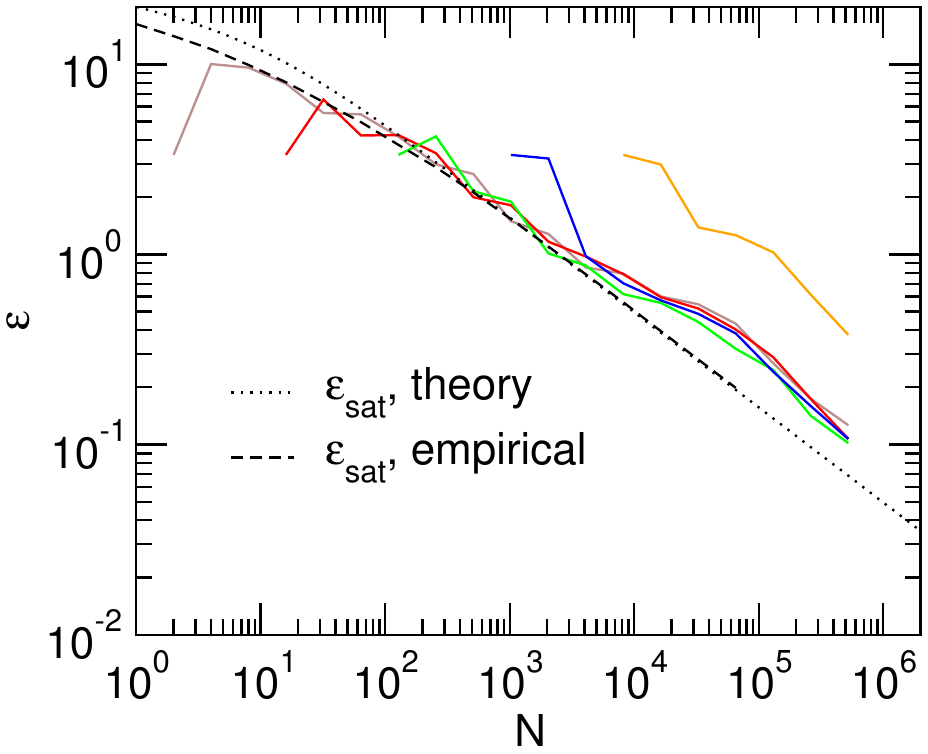}
	\caption{\label{fig:dfnt} Convergence behaviour for different initial effective number of samples $N$ for one-dimensional Brownian dynamics in a double-well potential (averaged over 32 runs). The curves are labeled by their $\eta^0$ value (equation \eqref{eq:eta}). Top: error $\varepsilon$ as a function of the collected number of samples $S= t/\Delta t_\Lambda$. 
	Bottom: $\varepsilon$ as a function of $N=N(t)$ compared with $\varepsilon_{\text{sat}}(N)$, obtained both empirically through simulation and theoretically using equation \eqref{eq:eps2satalln}.}
\end{figure}	
 We expect an analogous relation to be applicable for the error in $f(\lambda)$ in the reaction coordinate formulation of AWH and we obtain it by finding the approximate correspondences between a special case of the AWH method and metadynamics.
From equations \eqref{eq:omegalambdax}, \eqref{eq:df} and \eqref{eq:qkappa}  we see that assuming that $n_\Lambda=1$, $\rho$ uniform and further
that the bias $g(\lambda)$ and equilibrium distribution $\pi_0(x)$ are approximately constant within an umbrella width, the AWH $f$ update has the shape of a Gaussian function centered at $\xi$ of width 
$\sigma_\kappa = 1/\sqrt{\beta \kappa}$
and height $1/N\rho Z_\omega$:
\begin{align*}
|\Delta f(\lambda)| 
&\approx \ln\left(
1 + \frac{1}{N\rho Z_\omega}
e^{-\frac{1}{2\sigma_\kappa^2}
\left(\xi-\lambda\right)^2}
\right)\addnumber{\label{eq:df0allN}}\\
&\approx \frac{1}{N\rho Z_\omega}
e^{-\frac{1}{2\sigma_\kappa^2}
\left(\xi-\lambda\right)^2},
\end{align*}
where the second equation is obtained in the limit of large $N$.
We identify $\max_\lambda|\Delta f(\lambda)|$ in equation \eqref{eq:df0allN} 
with the dimensionless Gaussian height $\beta w$. By assuming $\rho$ uniform and further applying a Gaussian integral approximation, we obtain:   
$\rho Z_\omega \approx \upsilon := ({\sqrt{2\pi}\sigma_\kappa}/{L})^d$,
where $\upsilon$ is the fraction of the Gaussian volume to the volume of $\Lambda$.
After making the obvious changes of variable names the approximate AWH version of equation \eqref{eq:eps2satmtd} becomes,
\begin{align*}
\varepsilon^2_\text{sat}
&= \frac{L^2\upsilon}{\Delta t_\Lambda D}
 \ln\left(1 + \frac{1}{N}
 \frac{1}{\upsilon}\right) 
\zeta(\sigma_\kappa/L)
\addnumber{\label{eq:eps2satalln}}\\
&\approx
\frac{L^2}{N \Delta t_\Lambda D} 
\zeta(\sigma_\kappa/L),
\addnumber{\label{eq:eps2sat}}
\end{align*}
where $\Delta t_\Lambda$ is the time in between $\lambda$ samples 
and equation \eqref{eq:eps2sat} is valid for large $N$.
The factor $N\Delta t_\Lambda$ represents the effective sampling time. For a fixed sampling time, decreasing $\Delta t_\Lambda$ yields proportionally more samples $N$, but does not decrease the error, since inter-sample correlations increase as well.
We note that, as is the case for equation \eqref{eq:eps2satmtd}, the above equations are strictly only valid in the limit of continuous (frequent) updates.

Solving
equation \eqref{eq:eps0epssatN0} for $N^0$ with the help of equation \eqref{eq:eps2sat} now yields
\begin{equation}
N^0 \lesssim \frac{L^2}{\varepsilon_0^{2} \Delta t_\Lambda D}
\zeta(\sigma_\kappa/L).
\label{eq:n0}
\end{equation}
In principle, this is a recipe for choosing $N^0$ given $\varepsilon_0^2$. In practice however, $D$ can be challenging to estimate \cite{Tian:2014ec} and furthermore $D$ might vary as a function of $\lambda$.
Nonetheless, assuming $D$ is roughly known so that equation \eqref{eq:n0} can be applied, $N^0$ can be estimated to an order of magnitude. 
This is often enough since the initial exponential bootstrapping of $N$ is quite effective in desensitizing the method to variations in $N^0$. 

We demonstrate the validity of our proposed method setup  
by studying the convergence rate of the error $\varepsilon$ for 
$N^0=2^{1+3i}$, $i=0, 1, \ldots, 4$
(as in figure \ref{fig:nt}).
 In figure \ref{fig:dfnt}, we plot $\varepsilon$ both as a function of the number of collected samples $S = t/\Delta t_\Lambda$ (top) and as a function of the effective number of samples $N$ (bottom),  We again use the simple double-well test case.
To connect the results to equation \eqref{eq:eps0epssatN0} we label each curve by its 
$\eta^0$ 
value (which increases with $N^0$).
For $\varepsilon(S)$, we see that  
lower values of $\eta^0$ yield slightly increased error for short times, but
indistinguishable convergence rates for longer times,
 for which  $\varepsilon(t)\sim S^{-1/2} \propto t^{-1/2}$, as expected. 
The curve with the largest $\eta^0$ "separates" from the rest and displays an increased error even for long times. 
This shows that 
$\eta^0\lesssim 1$  is a good guideline for choosing $N^0$, while 
  $\eta^0 \gtrsim 10$ risks suboptimal convergence.

We now wish to find out  
how closely $\varepsilon(N)$ follows $\varepsilon_{\text{sat}}(N)$ for different values of $\eta^0$
and how this relates to the observed convergence in $\varepsilon(S)$.
We have plotted  both an empirical  $\varepsilon_{\text{sat}}$, obtained simply by setting $N\equiv N^0$ and waiting for the error to saturate, as well as the theoretical  $\varepsilon_{\text{sat}}$ obtained using equation \eqref{eq:eps2satalln}.
We see that all curves starting close to or below the saturated error curve relax to more or less the same $\eta(t)\gtrsim 1$, while the deviating, largest $\eta^0$ curve clearly lags behind. 
Obviously, the longer the simulation time, the less critical the choice of $N^0$ will become. For complex systems that are difficult to fully converge however, $N^0$ can substantially influence the final accuracy.

\section{\label{sec:applications} Applications}
Here we demonstrate the setup and illustrate the advantages of the AWH method for atomistic MD simulations. We calculate the PMF for two test cases: lithium acetate (LiAc) (section \ref{sec:liac}) and the 10-residue $\beta$-hairpin chignolin \cite{Honda:2004ca} (section \ref{sec:chig}).

All simulations were performed using a modified version of GROMACS 4.6\, \cite{Pronk:2013ef}. The reaction coordinate case of the AWH method was implemented as a module of the non-equilibrium pull code. 
Using the already existing replica exchange framework in GROMACS,  parallelization (see section \ref{sec:basicalgorithm}) was straightforward. We do not present results from multireplica simulations in this paper, however.

\subsection{\label{sec:awhsetup} Accelerated weight histogram setup}
Below we provide general guidelines for setting the input for an AWH simulation in PMF calculations. As we will see, many method parameters can take on default values. 

For instance,
the target distribution $\rho$ would most often be chosen uniform in the (estimated) region of interest $\Lambda$, possibly with a free energy cutoff of, say, $15$ $k_BT$.

 In addition, we update $f$ with every collected $\lambda$ sample, $n_\Lambda=1$. Single-sample updates which do not allow for any relaxation time
 might seem inconsistent with the fundamental assumption of the method that samples are generated from the current equilibrium distribution. 
In practice however, because of the adaptive biasing, the system  will in any case initially be far from relaxed. 
Moreover, as $N$ grows and $\Delta f$ shrinks, it is clear that the value of $n_\Lambda$ should matter less and less, since for large $N$ the logarithmic update in equation \eqref{eq:df} linearizes. More importantly, we have not been able to observe any measurable advantages to $n_\Lambda>1$ in our simulations (for which the computational effort of the AWH update step is negligible in comparison with the MD steps).

We also use a generic initial phase covering criterion (discussed in section \ref{sec:updatingN}).
To minimize any dependence on the point spacing, we use a temporary weight histogram $\Omega_N(\lambda)$ containing all the transition weights $\omega(\lambda)$ sampled at the current constant $N$ stage.
In the one-dimensional case, we double $N$ after both endpoints of $\Lambda$ have collected  
the weight corresponding to the peak of a Gaussian distribution of width $\sigma_\kappa$, i.e. 
$\Omega_N(\lambda_{\text{end}}) \geq \omega_{\text{peak}} = {\Delta \lambda}/{\sqrt{2\pi}\sigma_{\kappa}}$.
If a target distribution with free energy cutoff is used (equation \eqref{eq:pic}), we simply ignore the points falling outside of the cutoff  when checking if the criterion is fulfilled.
We straightforwardly generalize this  to the multidimensional case, $d>1$, by projecting the weight histogram onto each dimension,
and requiring analogously to the $d=1$ case that both endpoints of each one-dimensional interval have gathered the weight of the $d$-dimensional Gaussian peak,
$\omega_{\text{peak}} = \prod_i^d{\Delta \lambda_i}/{\sqrt{2\pi}\sigma_{\kappa_i}}$.
Obviously, in the multidimensional case this criterion does not guarantee that all relevant regions of $\Lambda$ have actually been covered. Still, it does ensure that \emph{some} extended path in $\Lambda$ has been explored.
As for the exit criterion of the initial phase, we use 
$N\geq N_{\text{ref}}$, one of the two similar criteria we proposed in section \ref{sec:updatingN}.


There are only three parameters that require more system specific attention: the force constant $\kappa$, the time interval between $\lambda$ updates $\Delta t_\Lambda$, and the initial effective number of samples $N^0$.
\begin{enumerate}
\item
$\kappa$, which couples $\xi$ to $\lambda$, is not particularly critical as long as the umbrella potential dominates that of the underlying free energy landscape.  
We give numerical examples in sections \ref{sec:liac} and \ref{sec:chig}.
Once $\kappa$ is set, the $\lambda$ point density is automatically determined as a function of the umbrella width 
$\sigma_\kappa = 1/\sqrt{\beta\kappa}$ in order to make transitions between points probable.
In our simulations we fix it to $\sim3\text{ points}/\sigma_\kappa$ per dimension.
We simply set the number of $\xi$ bins used for the deconvolution  equal to the number of $\lambda$ points.
\item
$\Delta t_{\Lambda}$, should be set as small as possible to minimize discontinuities, but still at least an order of magnitude larger than the MD time step to avoid introducing integration errors. For many biomolecules $\Delta t_{\Lambda} = 1$ ps could be used as a default value. See sections \ref{sec:liac} and \ref{sec:chig} for examples.
 In addition, $\Delta t_{\lambda}$ should be smaller than the diffusion time across an umbrella width in order to ensure that the dynamics of $\lambda$ does not slow down diffusion in $\xi$.
 Since diffusion anyhow is often slow along reaction coordinates, this is not a major constraint.
\item
$N^0$ has already been thoroughly discussed in section \ref{sec:n}. If  (an upper bound to) the diffusion coefficient $D$ can be estimated or there is previous experience from simulating similar systems, equation \eqref{eq:n0} can be used to estimate $N^0$. Alternatively, $N^0$ can by trial-and-error be set small enough to observe transitions in a shorter test run, but still large enough so that the observed variations in $f$ are of comparable magnitude to the expected barrier heights. Because of the initial exponential growth of $N$, the method is quite robust with respect to $N^0$ as long as it is not chosen too large.
\end{enumerate}

\subsection{\label{sec:mdsetup} Details of molecular dynamics setup}
Molecular dynamics simulations were performed with GROMACS 4.6.
The temperature was maintained at 298 K for LiAc and 300 K for chignolin using the v-rescale thermostat \cite{Bussi:2008cs}. Pressure was kept at 1 bar using Berendsen pressure coupling \cite{Berendsen:1984us}. Long-range electrostatics were calculated using Particle-Mesh Ewald \cite{Essmann:1995vj}. All bonds were constrained using the LINCS algorithm \cite{Hess:2008fl}. The time step was 2 fs for LiAc and 4 fs for chignolin (using virtual sites).
The force field used for LiAc was OPLS united atom \cite{Jorgensen:1996bs} with a modification using Kirkwood-Buff integrals to reproduce the activity \cite{Hess:2009du}.
For chignolin the AMBER99SB all-atom forcefield \cite{Hornak:2006gx} was used. Both systems were solvated in SPC/E water \cite{Berendsen:1987uu}; 1000 water molecules were added to LiAc and 2000 molecules to  chignolin. Two Na$^+$ ions were added to the solution of chignolin to neutralize the system.

\subsection{\label{sec:liac} Lithium acetate}
As a first atomistic application we study lithium acetate (LiAc) in water
and determine the PMF along the distance between a lithium ion and the carbonyl carbon of acetate, $\xi = d_{LiC}$. Ion pairing is a good test case, 
as for a contact ion pair to form, the solvation shells need to be rearranged, 
which requires conformational changes. 
Especially for small cations that bind water strongly, this leads to a high free energy barrier.
Furthermore, there is a narrow, small minimum at a very small distance (which does not appear if the distance to an oxygen is chosen as reaction coordinate). This narrow minimum is a test for the resolution and deconvolution of the method. 
\begin{figure}[tbp!]
\includegraphics[width=0.4\textwidth]{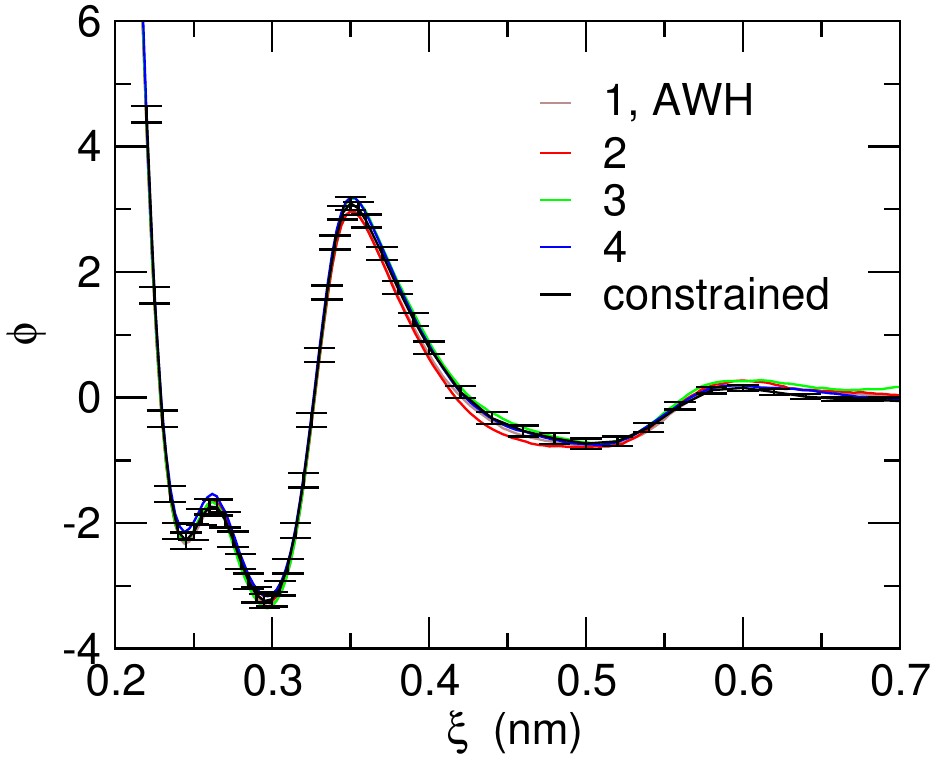}
\caption{\label{fig:liacpmf} PMF for LiAc for 4 independent, 50 ns long AWH runs and for the reference constraint calculation, based on 245 ns of simulation time.}
\end{figure}
We use a flat target distribution $\rho(\lambda)$ for $\lambda \in \Lambda =[0.2,0.7]$ nm.
 We expect the  LiAc system to be quite fast, so we set $\Delta t_\Lambda=$ 0.1 ps (50 time steps).
We empirically picked $N^0=2^{10}$ by setting it large enough to have reasonable fluctuations in $f$ after one $\Lambda$ covering.
 The force constant was initially set to $\kappa= 16\cdot 10^3$ kJ$\,$mol$^{-1}$nm$^{-2}$ ($\sigma_\kappa \approx 1\cdot 10^{-3}$ nm). After running a simulation for a short time we
 inspected the $\xi$ and $\lambda$ distributions and noticed that, 
 while the $\lambda$  distribution was relatively flat, $\xi\approx 0.35$ nm was only getting  0.16 of the mean number of samples, suggesting highly (negatively) curved free energy in that region. We therefore set $\kappa= 32\cdot 10^3$ kJ$\,$mol$^{-1}$nm$^{-2}$ ($\sigma_\kappa \approx 9\cdot 10^{-4}$ nm). This increased the sampling fraction to 0.56.

As a reference, we also calculated the PMF by constraining the distance $\xi$ and integrating the average constraint force. To resolve the steep barriers around the first minimum, we used 48 non-uniformly spaced distances and a total sampling time of 245 ns. 
We could also have used umbrella sampling which would have required not only non-uniform point spacing but also non-uniform umbrella widths. This demonstrates one of the main advantages of the AWH method:
we can globally make our umbrellas very narrow without increasing the computational cost, as opposed to umbrella sampling, where relaxation along the coordinates perpendicular to $\lambda$ is required for each $\lambda$ point individually.

In figure \ref{fig:liacpmf} we show the estimated PMFs at $t=50$ ns for 4 independent runs, together with the calculated reference curve.
We have subtracted the entropic term $-\ln(4\pi\xi^2)$ from the AWH profiles in order to exclude the effect of the  available phase space increasing  with $\xi$.
The standard deviation of the free energy difference between $\xi=0.3$ nm and at $\xi=0.7$ nm is 0.16 $k_BT$ for these AWH runs. For the constraint PMF, the estimated error is 0.27 $k_BT$ after 50 ns of simulation time, showing that the AWH method is at least as accurate as the method of constraints.

The dynamics during the final $1/t$ stage was compared to that of
 a regular MD simulation which had been biased with the PMF to obtain a flat free energy profile. In both cases the rate of the slowest process, the crossing of the barrier at 0.35 nm, was 4 times per ns. This shows that the AWH method does not slow down the kinetics, provided that $\Delta t_\Lambda$ is chosen sufficiently small.

We estimate the diffusion constant to 
$D\approx1.5\cdot10^{-4}$ $\text{nm}^2\text{ps}^{-1}$ for nearly the whole interval by looking at the mean square displacement of $\lambda$ over 5 to 50 ps for the later, diffusion-like, stages of an AWH run. 
This allows us to estimate $\varepsilon_{\text{sat}}$ from equation \eqref{eq:eps2satalln}, or,
using equation \eqref{eq:n0}, a reasonable value  for $N^0$.
  We obtain, using the estimated $\varepsilon^0 \approx 1.5$,
 $N^0 \approx 1200 \approx 2^{10} $. That is, our $N^0 =2^{10}=1024$ corresponds to $\eta^0$ close to 1.

 To further study the influence of $N^0$ on the convergence we extended our simulations to a range of $N^0$ values, 
$N^0=2^{4+3i}$, where $i=0, 1, \ldots, 4$. These correspond, respectively, to 
$\eta^0 \approx$ 0.1, 0.3, 0.9, 2.7, and 7.5.
We observed that all simulations displayed very similar convergence behavior, except those for which  $\eta^0\approx 7.5$ which was initially exploring slowly and displayed larger average error also for long times.

\subsection{\label{sec:chig} Chignolin}
We next explored a two-dimensional free energy landscape for the 10-residue $\beta$-hairpin chignolin\cite{Honda:2004ca}, in explicit water.
This is a more complex application than LiAc and is interesting to study because it contains features that also appear in conformational changes in larger proteins. In particular the formation of native hydrogen bonds can be difficult to sample, since a state with a large number of conformations needs to transition to the single, correctly hydrogen bonded conformation. 

\begin{figure}[tbp!]
\includegraphics[width=0.4\textwidth]{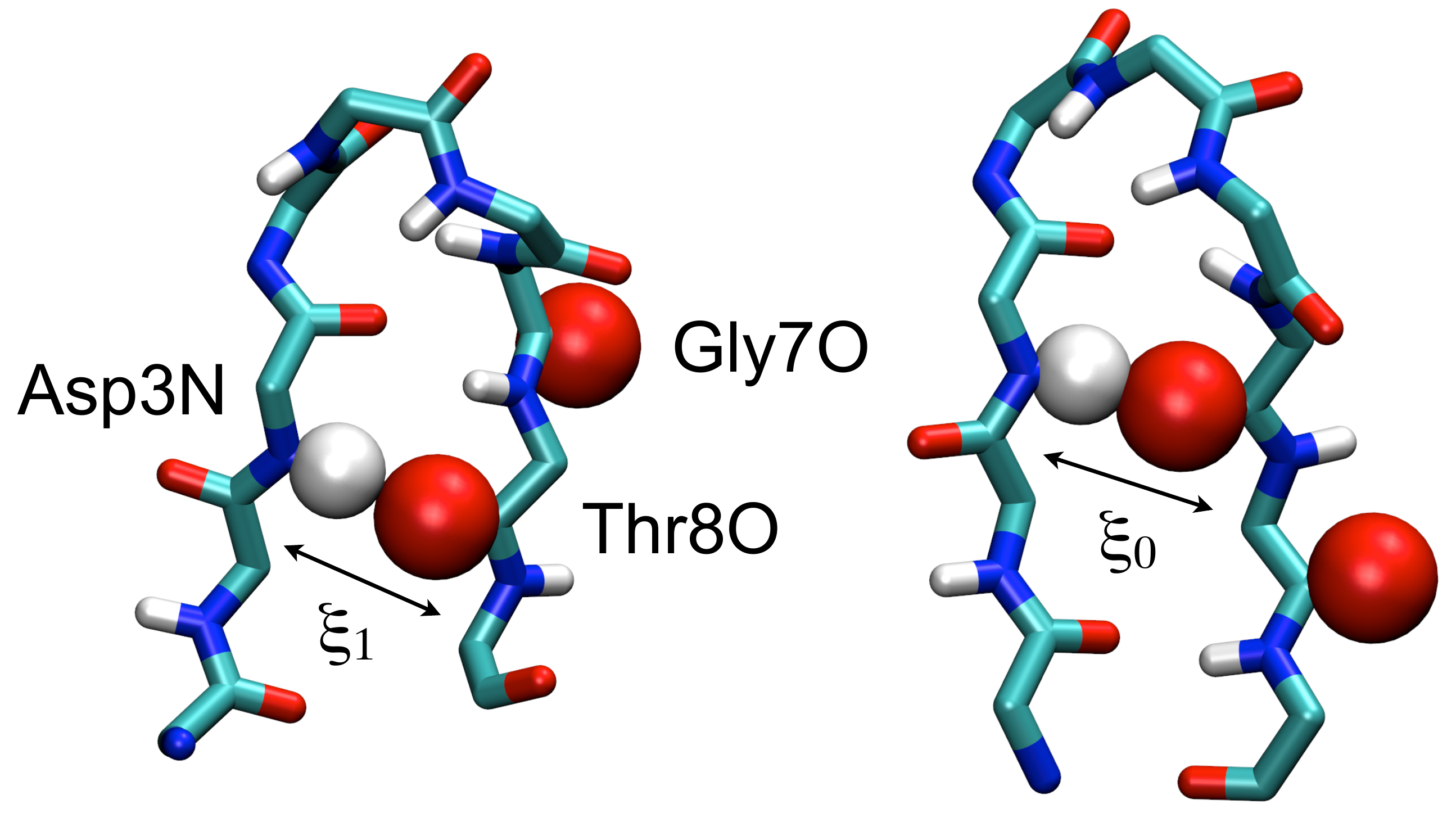}
\caption{\label{fig:chigconf} Folded conformations of chignolin and the two-dimensional reaction coordinate $(\xi_0,\xi_1)$, 
 the 
Asp3N-Gly7O and Asp3N-Thr8O distances, respectively.
 Left: native fold. Right: misfolded state.}
\end{figure}

Previous simulations of chignolin have shown \cite{Satoh:2006im} that, besides the native fold, there is a highly populated misfolded state. In this misfolded state, the outer Asp3N-Thr8O hydrogen bond in the native fold has been swapped to
a Asp3N-Gly7O hydrogen bond leading to a more tightly turning structure, see figure \ref{fig:chigconf}.
In order to map out the free-energy landscape between folded, misfolded and unfolded states, we define a two-dimensional reaction coordinate 
$\xi=(\xi_0,\xi_1)$, 
where $\xi_0$ and $\xi_1$ are the 
Asp3N-Gly7O and Asp3N-Thr8O distances, respectively.

To avoid sampling unphysical states, we use a target distribution $\rho$ with free energy cutoff $C=15$ (see equation \eqref{eq:pic}). We use uniform $\rho_0(\lambda)$, although an alternative could have been $\rho_0(\lambda)\sim1/|\lambda|^p$, for some $p>0$ to sample less of the unfolded configurations.
In this case using the covering criterion described in section \ref{sec:awhsetup}
seems reasonable since  the effective number of samples $N$ will likely not get doubled until both bonds have separately gone from opened to closed, or vice versa. 
Next, we set $\Lambda =[0.25,1]\times[0.25,1]\text{ nm}^2$ based on the fact that in the initial configuration, $\xi_1 \approx 0.3$  nm (closed) and $\xi_0\approx 0.6$ nm (open).
Further, $\kappa = 4\cdot 10^3$ kJ/mol$\cdot\text{nm}^2$ ($\sigma_\kappa = 0.02$ nm) for each dimension. 
We estimated $\Delta t_\Lambda$ = 1 ps (250 time steps), which is roughly the velocity decorrelation time for biomolecular systems, to be sufficiently small.  
We choose $N^0=2^{11}$ based on observations that this gives fast transitions without extreme free energy estimate fluctuations (alternatively, one could apply equation \eqref{eq:n0} using e.g.\ a rough square double well model of the landscape and an estimated upper bound for the diffusion coefficient).  

\begin{figure}[tbp!]
		\includegraphics[width=0.4\textwidth]{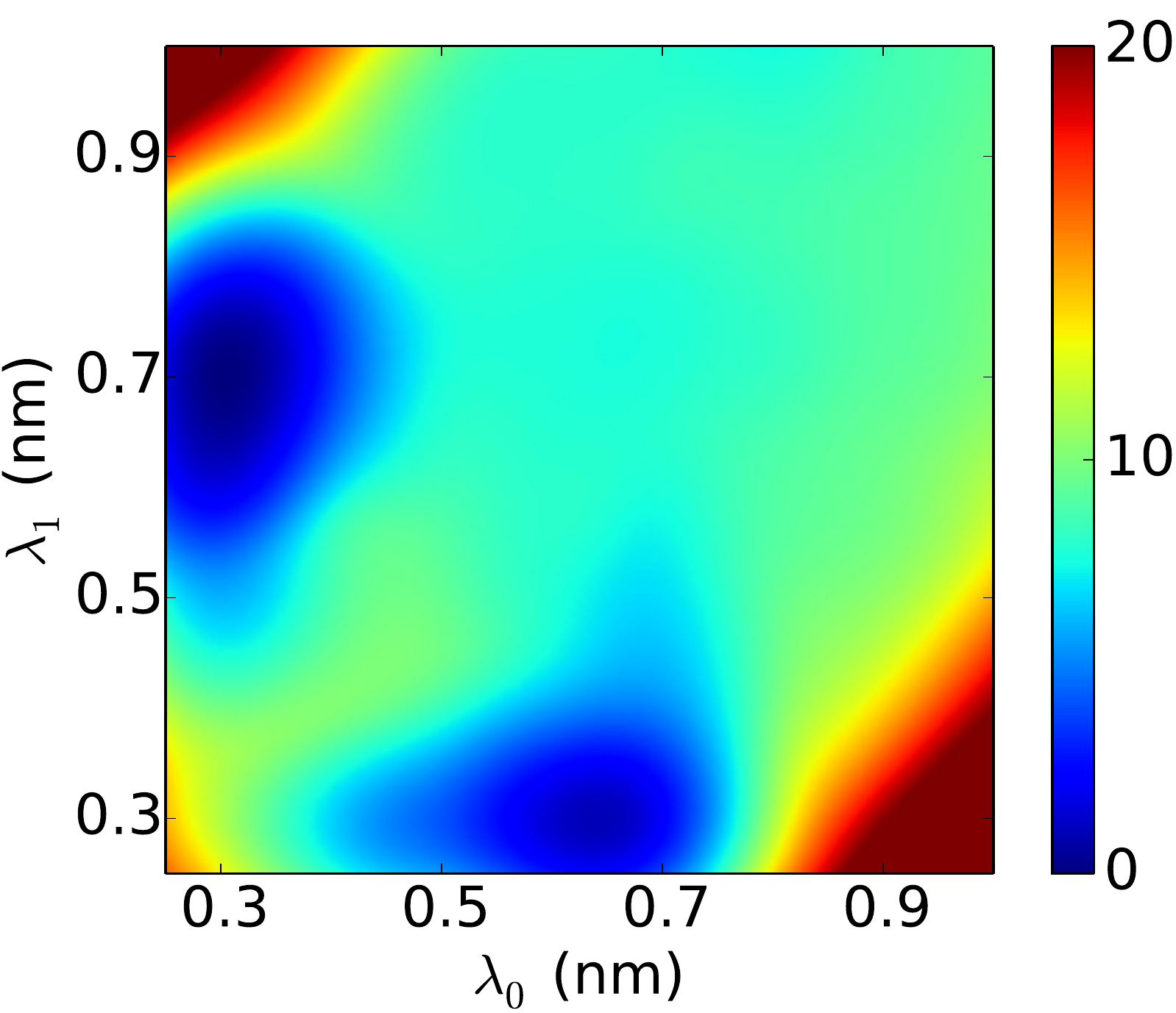}
		\includegraphics[width=0.4\textwidth]{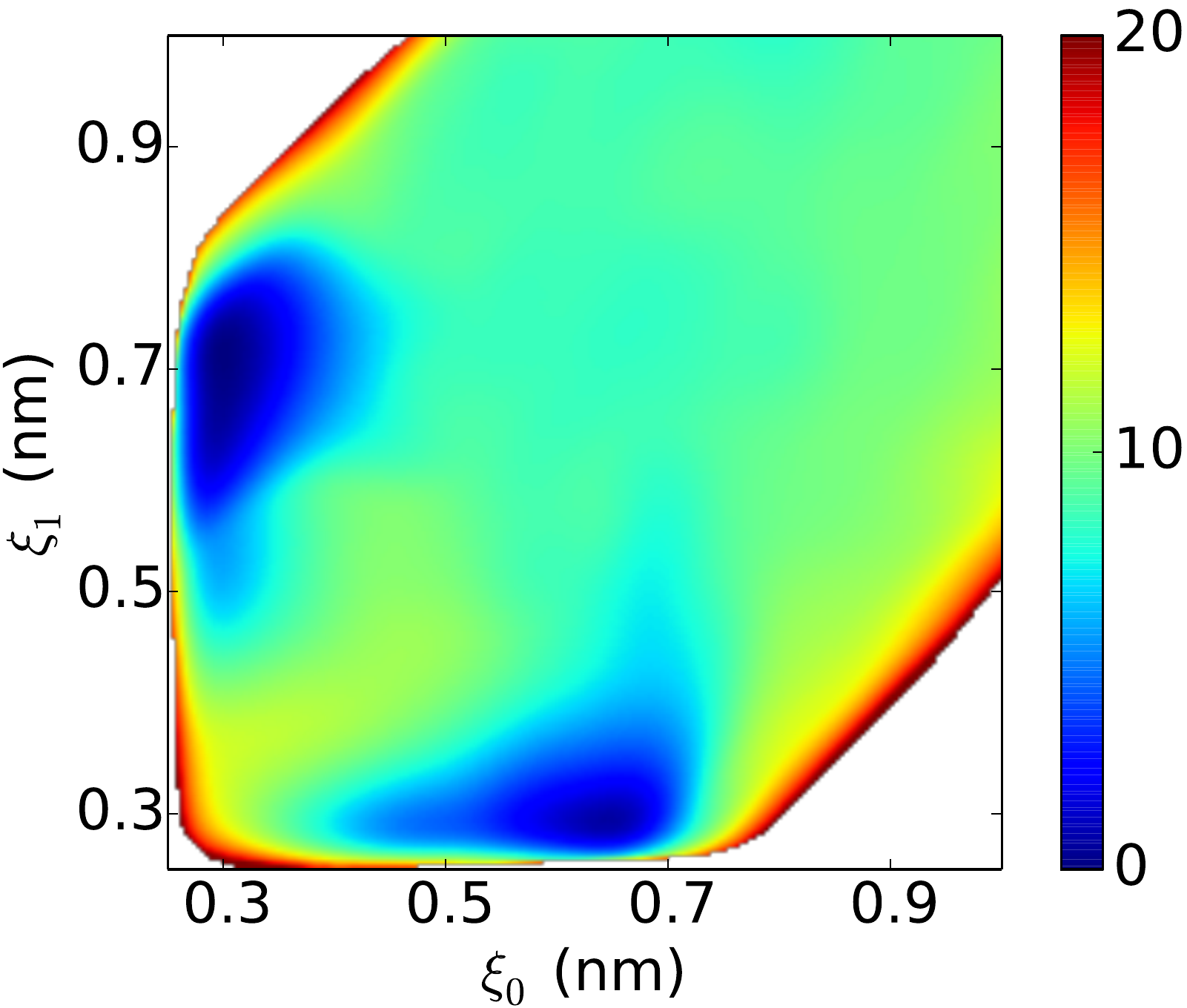}
	\caption{\label{fig:chigfs} The free energy estimates for chignolin at $t=1.2\,\mu$, averaged over 8 independent runs. The target distribution is uniform with a free energy cutoff of 15 $k_BT$. Top: $f(\lambda)$, the estimated convoluted free energy. Bottom: the PMF $\phi(\xi)$, the deconvolution of $f(\lambda)$.}
\end{figure}	

Figure \ref{fig:chigfs} shows the final, $t=$ 1.2 $\mu$s, free energy $f(\lambda)$ (top) and PMF $\phi(\xi)$ (bottom) averaged over  8 independent replicas.
The figure further illustrates a major advantage of using a target distribution with a cutoff on the free energy: the otherwise critical choice of $\Lambda$ is eliminated.
 Without such a cutoff one would have to carefully set up boundaries
that include all important states, but fence off uninteresting ones. 
Including improbable states does not only risk inefficient sampling but can also lead to instabilities due to large biasing forces.

\begin{figure}[tbp!]
\includegraphics[width=0.4\textwidth]{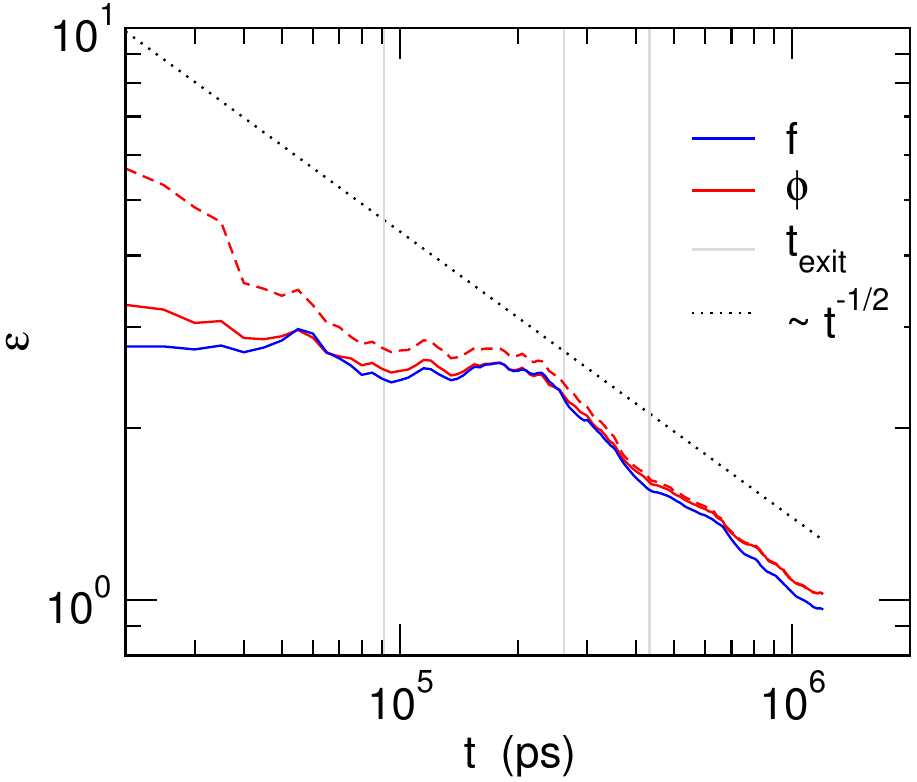}
\caption{\label{fig:chigsdfs} Convergence of the error in $f(\lambda)$, the convoluted free energy, and $\phi(\xi)$, the PMF, for chignolin.  The corresponding dashed curve for $\phi$ is the error divided by the fraction of explored space at time t relative to the final time. 
The average exit time $t_{\text{exit}}$, together with its one standard deviation error bars, are given by
vertical lines.}
\end{figure}

We find that the minimas corresponding to the two folded states are comparably populated, with a preference for the misfolded state.  Within the current accuracy  on the order of $\sim 1$ $k_BT$ (figure \ref{fig:chigsdfs}), our results are consistent with previous work\cite{Kuhrova:2012wt} where the native and misfolded state were found to be approximately equally populated.
We note that a different water model was used in \cite{Kuhrova:2012wt}.

The error in the convoluted free energy $f(\lambda)$ and the PMF $\phi(\xi)$ follow each other closely, as can be seen in figure \ref{fig:chigsdfs}. 
Obviously, because of the sampling-based deconvolution, the PMF error can only be calculated in a domain which has already been explored by all simulations. The exclusion of yet to be explored regions generally leads to an initial underestimation of the error. 
To separate this effect, we have included the dashed curve in figure \ref{fig:chigsdfs}, for which the error in $\phi$ has been divided by the fraction of included space at time $t$, relative to the final time.
In the initial stage, $t<t_{\text{exit}}$, where $t_{\text{exit}}$ is the mean exit time,  $\varepsilon(t)$  is characterized by a relatively flat plateau in which exploration of new areas is taking place simultaneously as the error is being reduced in already visited regions. 
In the final stage, the two $\phi$ curves are basically the same and $\varepsilon\sim t^{-1/2}$ convergence is recovered.

Our choice of a flat target distribution targets uniform accuracy in the entire explored reaction coordinate space, including relatively high free energy regions where the peptide is unfolded. 
If one is willing to accept significantly increased errors outside of the folded regions, a parameter extension in temperature space (i.e. $\lambda=T$) could be an alternative.
Using temperature replica exchange\cite{Sugita:1999cl} would also be possible since chignolin is a relatively small system.
For larger systems however, the applicability of replica exchange methods are severely limited by the large number of replicas needed\cite{Fukunishi:2002hb}.

We post-validated our choice of $N^0$ by estimating the diffusion constant. Using the mean square displacement method we obtained $D\approx 5\cdot 10^{-5}$  $\text{nm}^2\text{ps}^{-1}$ in the two minimas and $D\approx 1\mbox{--}2 \cdot 10^{-6}$  $\text{nm}^2\text{ps}^{-1}$  in the unfolded region, for starting times $t\geq 300$ ns and  time intervals 100 to 500 ps. Together with our estimation $\varepsilon^0 \approx 3$, equation \eqref{eq:eps2satalln} implies $\eta^0 \approx 0.7$ for the simulations. This shows that our trial-and error choice of $N^0$  is consistent with $\eta^0\lesssim 1$.

\section{\label{sec:conclusion} Conclusion}
The AWH method is a flexible and efficient adaptive biasing method for free energy calculations.
Its very general extended ensemble formulation opens up for numerous practical extensions, of which we have demonstrated two: a full resolution calculation procedure of the potential of mean force, and the use of a free energy dependent target distribution, which can be extremely helpful in avoiding sampling of improbable states. 
Furthermore, the AWH method is straightforward to implement, both in its serial and parallel version. We intend to include it in version 6.0 of the GROMACS molecular simulation package.

A core feature of the AWH method is the use of a weight histogram. This allows for efficient utilization of the transition history in order parameter space, both in the free energy updates and as a means of monitoring the simulation. 
Moreover, as a result of using a weight histogram  
there is no tedious discretization or binning procedure involved in setting up an AWH simulation, and the method efficiency does not depend critically on the point density.
An additional major advantage of the AWH method is that repeated passes are made over the reaction coordinate, which allows for multiple pathways. 
The quality of the initial starting structure is thus not critical, since the system will have time to relax during the course of the simulation. For the same reason, initially choosing a large update size is not a problem even though this may initially push the system into strained configurations.

 The most important factor in PMF calculations remains the, often difficult, choice of reaction coordinate. Nonetheless, for a good reaction coordinate, the AWH method makes the calculation of PMFs straightforward. 
The AWH method is furthermore helpful as an aid for detecting a bad choice of reaction coordinate, since the weight histogram will deviate significantly from the target distribution if there are issues. The early exploration of reaction coordinate space is usually fast and can be used to assess the current choice of reaction coordinate.

From a practical point of view, we have provided concrete guidelines and numerically illustrated how to customize and initialize AWH simulations for PMF calculations of molecular systems. This knowledge will be of great help in our future simulations of more complex systems.

\section*{Acknowledgments}
This work was supported by the European Research Council
(grant nr. 258980) and the Swedish e-Science Research Center.
Computer resources were provided through the Swedish
National Infrastructure for Computing (SNIC 001/12-280).

\appendix*
\section{\label{app:efftempdistr} Effective temperature target distribution}
In this section we describe in detail how the AWH method can be set up to obtain an algorithm very similar to  well-tempered metadynamics\cite{Barducci:2008ua}.

In well-tempered metadynamics, the rate of change of the biasing potential $V(\xi,t)$ is given by 
(adding a tilde to the notation in \cite{Barducci:2008ua} whenever necessary to minimize confusion with AWH variables):
\begin{equation}
\dot{V}(\xi,t) = 
\frac{\tilde{\omega}\Delta T}{\Delta T + \tilde{\omega}\tilde{N}(\xi,t)} \delta_{\xi,\xi(t)}
,\label{eq:wtmetadynamics}
\end{equation}
where  dot denotes time derivative, $\tilde{\omega}$ is the energy deposit rate, $\Delta T$ is the effective temperature increase and
 $\tilde{N}(\xi,t) = \integ_0^t \delta_{\xi,\xi(t')}\diff t'$ is the histogram of $\xi$.  
 In practice the deposit represented by $\delta_{\xi,\xi(t)}$ is replaced by (e.g.) a finite width Gaussian. The initial Gaussian height is from equation \eqref{eq:wtmetadynamics} given by $\tilde{\omega}\tau_G$, where $\tau_G$ is the time interval in between deposits. As $\tilde{N}$ grows, we see that the height decreases as $1/\tilde{N}(\xi,t)$. 
 With this biasing procedure the distribution along $\xi$ converges to 
  $P_{\infty}(\xi) = e^{-\frac{T}{T+\Delta T}\Phi(\xi)}/Z$, 
  i.e.\, a Boltzmann distribution for an effective temperature $T+\Delta T$.
%
%

Within the AWH formalism we can obtain a analogous update scheme by a special choice of the target distribution $\rho$ that explicitly depends on the sampling history.
The optimal choice of $\rho$ is generally a trade-off between exploring new regions (repulsion) and improving sampling in familiar, high-probability regions (attraction). It is clear from equation \eqref{eq:gopt} that the  free energy update can be used to  bias the future simulation either by transferring $\Delta f$ to the tuning factor $g$, effectively giving rise to a repulsive force, or to $\ln\rho$, yielding an attractive force, or both. 
We can parametrize this division by introducing a "stickiness" factor $0\leq b\leq 1$ representing the fraction of $\Delta f$ contributing to $\Delta \ln\rho$, which sets the tendency to stick to already visited regions. 
Here we ignore the cases $b>1$ and $b<0$ which correspond to an effectively decreased temperature and negative effective temperature, respectively.

 Explicitly, we let the reference weight histogram $W=N\rho$ grow
by adding the accumulated sum of transition weights, scaled by $b$, on top of it. That is, 
\begin{equation*}
W(\lambda,t) = W^0(\lambda) + b\Omega(\lambda,t),
\end{equation*}
where $W^0(\lambda) = N^0\rho^0(\lambda)$ determines the initial conditions. 
Consequently, the growth of $N(t) = \sum_\lambda W(\lambda,t)$ is still linear, but scaled by $b$.
We assume for simplicity that $n_\Lambda = 1$ and see that this choice implies an update for $\rho$, $\ln\rho\leftarrow \ln\rho + \Delta \ln \rho$, given by (disregarding irrelevant constants)
\begin{align*}
\Delta \ln \rho(\lambda,t) 
&= 
\ln\left(\frac{W(\lambda,t)+ b\omega(\lambda,t)}{N(t) + b}\right)
- \ln\left(\frac{W(\lambda,t)}{N(t)}\right)\\
&= 
\ln\left(1 + b\frac{\omega(\lambda,t)}{W(\lambda,t)}\right)\\
&\approx
b\frac{\omega(\lambda,t)}{W(\lambda,t)},\addnumber{\label{eq:dlnrhow}} 
\end{align*}
where the approximation is valid for large $N$. The same approximation into equation \eqref{eq:df}  leads to 
$\Delta f(\lambda,t) \approx -\frac{\omega(\lambda,t)}{W(\lambda,t)}$. Maintaining equation \eqref{eq:gopt} intact thus necessitates
\begin{align*}
\Delta g(\lambda,t)
&\approx 
-(1-b)\frac{\omega(\lambda,t)}{W(\lambda,t)} \\
&= 
-\omega(\lambda,t) \left(\frac{1-b}{W^0(\lambda,t) + b\Omega(\lambda,t)}\right).
\addnumber{\label{eq:dgw}}
\end{align*}
For the case $b=0$ (non-stick), $W$ has no memory,   
leading to $\Delta\ln\rho=0$, $\Delta f = \Delta g$.
For $b=1$ (sticky) on the other hand $W$ has "perfect" memory and $\Delta \ln\rho = -\Delta f$, $\Delta g = 0$ (no bias).
These cases, respectively, correspond to the well-tempered metadynamics cases of $\Delta T\to\infty$, $P_\infty$ uniform and $\Delta T = 0$, $P_\infty = e^{-\Phi}/Z$.

We see from from equation \eqref{eq:dgw} that $g$, just as $V$, indeed acts repulsively in the sense that $\Delta g$ becomes increasingly negative in regions where $\omega$ peaks, which according to equation \eqref{eq:omegalambdax} decreases the probability of returning to that region in future transitions. In the case of $V$ however, $\dot{V}$ becomes increasingly \emph{positive} along the trajectory of the reaction coordinate.

To continue with this comparison, we note that in the AWH method, setting $W^0$ fixes the initial magnitude of $\Delta f$, while for well-tempered metadynamics the initial conditions are specified by the deposit rate $\tilde{\omega}$, fixing the size of the  initial bias update.
To clarify the connection between $\Delta g$  and that of equation \eqref{eq:wtmetadynamics} the initial conditions must be chosen consistently for both methods.  This is achieved by scaling $\Delta g$ in equation \eqref{eq:dgw} as 
$W^0 = (1-b)\tilde{W}^0$ where now $\tilde{W}^0$ should be possible to relate to $\tilde{\omega}$.
Furthermore making the change of variables
$b = T/(T+\Delta T)$ ($\Delta T\geq 0$) 
and rearranging
we obtain
\begin{equation}
\Delta g(\lambda,t)
\approx 
-\frac{\frac{1}{\tilde{W}^0(\lambda)}\Delta T}
{\Delta T + \frac{1}{\tilde{W}^0(\lambda)} 
\Omega(\lambda,t) T}\omega(\lambda,t).
\label{eq:dgapprox}
\end{equation}
We define the unitless 
$\Delta\tilde{g}(\xi,t) = -\frac{\tau_G}{T} \dot{V}(\xi,t)$
for sake of comparison and 
obtain from equation \eqref{eq:wtmetadynamics},
\begin{equation}
\Delta \tilde{g}(\xi,t) 
=-\frac{\frac{\tilde{\omega}}{T}\Delta T}{\Delta T + \frac{\tilde{\omega}}{T}\tilde{N}(\xi,t)T} \delta_{\xi,\xi(t)}\tau_G
\end{equation}
which is of the same form as equation \eqref{eq:dgapprox} after the straightforward correspondences have been set up.


\bibliography{awh}{}

\end{document}